\documentclass[amssymb,prb,twocolumn,showpacs,superscriptaddress]{revtex4}
\usepackage{epsfig}
\usepackage{dcolumn}
\usepackage{amsmath}
\begin{document}

\title{Chirality in Coulomb-blockaded quantum dots}
\author{David S\'anchez}
\affiliation{Departament de F\'{\i}sica,
Universitat de les Illes Balears, E-07122 Palma de Mallorca, Spain}
\affiliation{D\'epartement de Physique Th\'eorique,
Universit\'e de Gen\`eve, CH-1211 Gen\`eve 4, Switzerland}
\author{Markus B\"uttiker}
\affiliation{D\'epartement de Physique Th\'eorique,
Universit\'e de Gen\`eve, CH-1211 Gen\`eve 4, Switzerland}
\date{\today}

\begin{abstract}
We investigate the two-terminal nonlinear conductance of a Coulomb-blockaded
quantum dot attached to chiral edge states. Reversal of the applied magnetic
field inverts the system chirality and leads to a different polarization
charge. As a result, the current--voltage characteristic is not
an even function of the magnetic field. 
We show that the corresponding magnetic-field asymmetry
arises from single-charge effects and vanishes in the limit
of high temperature.
\end{abstract}

% 73.23.-b -> Electronic transport in mesoscopic systems 
% 73.50.Fq -> High-field and nonlinear effects
% 73.63.Kv -> Quantum dots
\maketitle

{\em Introduction}.---The Onsager-Casimir symmetry relations~\cite{ons31,cas45} establish
that the linear-response transport is even under
reversal of an external magnetic field. It is then of fundamental interest 
to investigate the conditions under which one can see deviations from the Onsager
symmetries as one enters the nonlinear regime. 
Recently, it has been shown~\cite{san04,spi04} that in nonlinear mesoscopic
transport there arise magnetic-field asymmetries entirely
due to the effect of electron-electron interactions in the nonlinear
regime.\cite{but93}. These works have been focused on quantum dots
with large density of states and connected
to leads via highly conducting openings (typically, quantum point contacts
supporting more than one propagating mode).\cite{san04}
Recent experiments by Zumb\"uhl {\it et al.}\cite{unpub} on large chaotic 
cavities are in good agreement 
with theory. In non-linear transport magnetic field-asymmetries can 
occur under a wide variety of conditions.\cite{rik05,wei05}  
In particular, in our work, we considered\cite{san04}
a quasi-localized level separated from the leads
with tunnel barriers but neglected single-charge effects.
Therefore, it is natural to ask whether magnetic field asymmetries
are  visible in the
Coulomb-blockade regime.\cite{kou97,bee91,ale02} Since Coulomb energies 
can be much larger than the energy scales for quantum interference, magnetic-field
asymmetries induced by 
single electron effects should be visible at much higher temperatures. 

The electrostatic approach used in the classical model of
Coulomb blockade\cite{bee91} predicts a potential difference $U_d$ between the quantum
dot (QD) and the reservoirs which depends on the QD charge $Q_d$,
\begin{equation}\label{eq_ud}
\phi_d=\frac{Q_d}{C_\Sigma} + \phi_{\rm ext} \,,
\end{equation}
and on an external potential $\phi_{\rm ext}$ related to the polarization
charge $Q_{\rm ext}$ externally induced by nearby reservoirs and gates:
\begin{equation}\label{eq_ext}
\phi_{\rm ext}=\frac{Q_{\rm ext}}{C_\Sigma}=
\frac{\sum_\alpha C_\alpha V_\alpha}{C_\Sigma} \,,
\end{equation}
where the sum extends over all leads. This model assumes a uniform screening
potential described by the QD (geometric) capacitance couplings $C_\alpha$
with the contacts. The total capacitance of the equivalent circuit
is thus $C_\Sigma=\sum_\alpha C_\alpha$.
\begin{figure}[b]
\centerline{
\epsfig{file=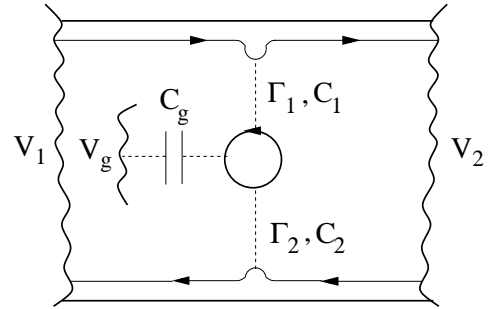,angle=0,width=0.35\textwidth,clip}
}
\caption{Sketch of the system under consideration. The antidot
is coupled to chiral edge states via tunnel barriers
acting as leaky capacitors. A back-gate
contact controls the dot occupation with a capacitive coupling $C_g$.
When the magnetic field is reversed, both edge states invert their
propagating direction.}
\label{antidot}
\end{figure}

Consider now a two-terminal sample in the quantum Hall regime
(see Fig.~\ref{antidot}) with one edge
state running along each side (top and bottom). With the help of gates
it is possible to create in the center a potential hill which behaves as a
tunable quasi-localized state coupled to edge states acting as source
and drain reservoirs. The resulting antidot~\cite{for94} connects the edge
states in two ways:\cite{san04,chr96}
(i) {\em scattering coupling}, in which electrons
tunnel from the edge states to the antidot, and (ii) {\em electrostatic coupling},
in which the antidot screening potential feels the repulsion through
capacitive couplings: $C_1$ ($C_2$) 
between the dot and the upper (lower) edge state.
The system has a definite chirality determined from the magnetic field
direction (upward or downward) since, e.g., the upper edge state
originates from the left terminal for a given field $+B$
but carries current from the right terminal for the opposite field
direction $-B$. Thus, the nonequilibrium polarization charge becomes
$Q_{\rm ext}(+B)=C_1 V_1+C_2 V_2$ and $Q_{\rm ext}(-B)=C_2 V_1+C_1 V_2$,
which is clearly magnetic-field asymmetric whenever the capacitance
coupling is asymmetric. Thus, we expect that the current traversing
the dot is not an even function of $B$.

The qualitative argument above can be traced back to the oddness of the
Hall potential.\cite{san04} We investigate now the effect in detail to give
precise predictions for the dependence of the magnetic-field asymmetry
on temperature, bias and gate voltages.

{\em Model}.---Electrons from lead $\alpha$ tunnel onto the dot via the edge
states with a transmission probability characterized
by a Breit-Wigner resonance with a width
$\Gamma_\alpha$. We assume that transport is governed by transitions between QD
ground states, which is a good approximation when both temperature and bias
voltages are much smaller than any excitation energy.
Then, the scattering matrix\cite{ale02,sta96} $S_{\alpha\beta}^N$ for the transition
from $Q_d=N-1$ electrons to $Q_d=N$ electrons when an electron is transmitted
from lead $\alpha$ to lead $\beta$,
\begin{equation}\label{eq_s}
S_{\alpha\beta}^N(E)=\delta_{\alpha\beta}-
i\frac{\sqrt{\Gamma_\alpha^N\Gamma_\beta^N}}
{E-\mu_d(N)+i\Gamma^N/2}\,,
\end{equation}
has a complex pole with a real part associated to the QD electrochemical
potential $\mu_d(N)$. The total resonance width is proportional to
$\Gamma^N=\sum_\alpha \Gamma_\alpha^N$. The widths fluctuate according to
the Porter-Thomas distribution but in what follows we neglect
intradot correlation effects in $\Gamma$ and take it as
energy independent, which works well provided bias variations
are much smaller than the barrier height.

We emphasize that the scattering matrix
is not only a function of the carrier's energy $E$ but also depends
on the full electrostatic configuration via $\mu(N)=E(N)-E(N-1)$, where
E(N) is the ground-state energy of a N-electron QD,
\begin{equation}\label{eq_etot}
E(N)=\sum_{i=1}^N \varepsilon_i + \frac{(Ne)^2}{2C_\Sigma}
-eN \sum_\alpha \frac{C_\alpha V_\alpha}{C_\Sigma}\,.
\end{equation}
In Eq.~(\ref{eq_etot}),
$E(N)$ consists of two terms. First, the kinetic energy
is a sum over QD single-particle levels
$E_k=\sum_{i=1}^N \varepsilon_i$ arising from confinement. These levels
may be, in general, renormalized due to coupling to the leads:
$\varepsilon_N \to \varepsilon_N + (\Gamma^N/\pi) \ln |(D-E)/(D+E)|$
with $D$ the bandwidth assuming flat density of states in the leads.
The renormalization term is a slowly increasing function
of $E$ and can be safely neglected. Therefore, the kinetic energy 
is invariant under reversal of $B$.
The second contribution to $E(N)$ is the potential energy $U(N)$
which depends on the charge state of the dot and the set
of applied voltages including nearby gates.
We assume that the dot is in the presence of a back-gate potential $V_g$
which controls the number of electrons at equilibrium via
the capacitance coupling $C_g$ (see Fig.~\ref{antidot}).
Then, the QD charge, which is
quantized to a value $Q_d=-Ne$ in the Coulomb valleys, determines the QD potential
from the discretized Poisson equation,
\begin{equation}\label{eq_poi}
C_1 (\phi_d-V_1)+C_2 (\phi_d-V_2)+C_g (\phi_d-V_g)=-Ne\,,
\end{equation}
which amounts to the Hartree approximation, disregarding exchange and pairing effects.
These effects might be important in certain situations\cite{ale02}
but we shall see below that this level of approximation already suffices
to obtain a sizable magnetic-field asymmetry.

Equations~(\ref{eq_ud}) and~(\ref{eq_ext}) are readily derived from
Eq.~(\ref{eq_poi}). Then, we find that the QD potential energy reads 
\begin{equation}\label{eq_u+b}
U(N,+B)=\frac{N^2\mathcal{U}}{2}-eN\left(
\frac{C_1}{C_\Sigma}V_1+\frac{C_2}{C_\Sigma}V_2+\frac{C_g}{C_\Sigma}V_g
\right)\,,
\end{equation}
where $C_\Sigma=C_1+C_2+C_g$ and $\mathcal{U}=e^2/C_\Sigma$.
We now reverse the magnetic field:
\begin{equation}\label{eq_u-b}
U(N,-B)=\frac{N^2\mathcal{U}}{2}-eN\left(
\frac{C_2}{C_\Sigma}V_1+\frac{C_1}{C_\Sigma}V_2+\frac{C_g}{C_\Sigma}V_g
\right)\,.
\end{equation}

From Eqs.~(\ref{eq_u+b}) and~(\ref{eq_u-b}) it is clear that the
QD electrochemical potential shows a magnetic-field {\em asymmetry},
$\Phi_\mu=[\mu(N,+B)-\mu(N,-B)]/2$, given by
\begin{equation}\label{eq_muas}
\Phi_\mu=\frac{C_2-C_1}{2C_\Sigma} (V_1-V_2)\,.
\end{equation}
Since $\mu(N)$ determines the position of the differential conductance
resonance, it follows that the $I$--$V$ characteristics of the antidot
is asymmetric under $B$ reversal.
We remark that this model assumes full screening of the charges
injected in the dot, i.e. the local potential neutralizes
the excess charge: $C_\alpha\ll e^2\nu_\alpha$ with
$\nu_\alpha$ the density of states of edge state $\alpha$. 
Deviations from this limit would probably decrease the size of
the asymmetry.\cite{san04}
Finally, we emphasize that magnetic field asymmetries develop
only to the extent that capacitive
interactions with surrounding contacts are considered.
\begin{figure}
\centerline{
\epsfig{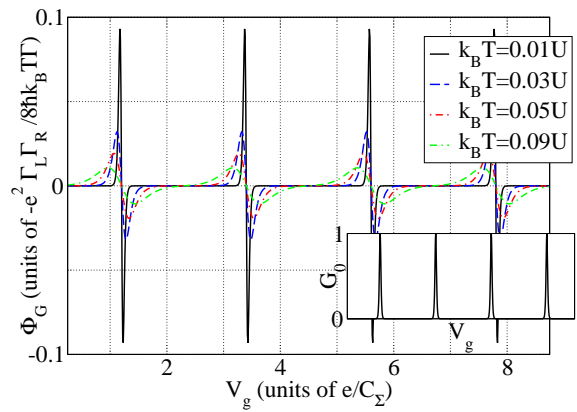}
}
\caption{(Color online). Magnetic-field asymmetry of the
differential conductance versus gate voltage for different
temperatures. We set $C_1+C_2=C_g=0.5$ ($C_\Sigma=1$),
asymmetry factor $\eta=0.5$, $\Delta=0.1\mathcal{U}$, $\Gamma=0.002\mathcal{U}$
and $V=0.005\mathcal{U}$ ($\mathcal{U}=e^2/C_\Sigma$).
Inset: Coulomb-blockade oscillations of the linear
conductance ($V=0$) for the same parameters and $k_B T=0.01\mathcal{U}$.}
\label{fig1}
\end{figure}

{\em Results}.---The current around the $N-1\to N$ resonance
for spinless electrons reads
\begin{equation}\label{eq_cur}
I_N (B)=-\frac{e}{h} \int dE\, {(S_{12}^N)}^\dagger
{S_{12}^N} [f_1(E)-f_2(E)]\,,
\end{equation}
where the scattering matrix $S$ from Eq.~(\ref{eq_s}) depends on $B$
because the QD potential response is asymmetric under $B$
reversal, as shown above. $f(E)$ is the Fermi function
and we take $V_1=-V_2=V/2$. Our goal is to calculate the asymmetry,
\begin{equation}\label{eq_condas}
\Phi_G=\frac{G_N(+B)-G_N(-B)}{2} \,,
\end{equation}
of the differential conductance $G_N=dI_N/dV$.

In the classical Coulomb-blockade regime, one neglects quantum fluctuations
in $Q_d$. Since the coupling to the leads causes a finite
lifetime of the QD charges, $Q_d$ is quantized only
when $k_B T\gg \Gamma^N$. Furthermore, one assumes that
there is no overlap between the distinct resonances,
thereby the mean level spacing in the dot $\Delta\varepsilon \gg \Gamma$.
Hence, we expand Eq.~(\ref{eq_cur}) to leading order in $\Gamma$
and obtain $G_N(V)$ for $B>0$:
\begin{equation}\label{eq_condB}
G_N(V,+B)=-\frac{e^2}{\hbar} \frac{\Gamma_L^N \Gamma_R^N}
{4C_\Sigma k_B T\Gamma^N}
[y_2(V)+y_1(-V)] \,,
\end{equation} 
with
\begin{equation}
y_\alpha(V)=(C_\alpha +C_g/2)
\cosh^{-2}\left(\frac{\tilde{\varepsilon}_N+
eV\frac{C_\alpha +C_g/2}{C_\Sigma}}{2k_BT}\right)
\,,
\end{equation} 
for $\alpha=1,2$ where $\tilde{\varepsilon}_N=\varepsilon_N-E_F+\mathcal{U}(N-1/2)-eC_g
V_g/{C_\Sigma}$ with $E_F$ the Fermi energy in the leads.
For $B<0$ one must make in Eq.~(\ref{eq_condB}) the replacement
$1 \to 2$ and $V \to -V$.
Then, our expression predicts a magnetic-field asymmetry
which arises only in the nonlinear conductance (for voltages $V\neq 0$)
and only due to electrostatic interactions with the leads.
For $V=0$ we reproduce the expression of the linear conductance
$G_0=G(V=0)$ as a function of $V_g$.\cite{bee91}
%\begin{equation}
%G_0=-\frac{e^2}{\hbar} \frac{\Gamma_L^N \Gamma_R^N}{4k_B T\Gamma^N}
%\frac{1}{\cosh^{2}(\tilde{\varepsilon}_N/2k_BT)}\,,
%\end{equation}
$G_0$ is independent on the sign of $B$, thus fulfilling the Onsager relation.
Sharp Coulomb-blockade peaks are observed in the oscillating
$G_0$ as a function of $V_g$ when $k_B T\ll e^2/C_\Sigma$
(see inset of Fig.~\ref{fig1}).

\begin{figure}
\centerline{
\epsfig{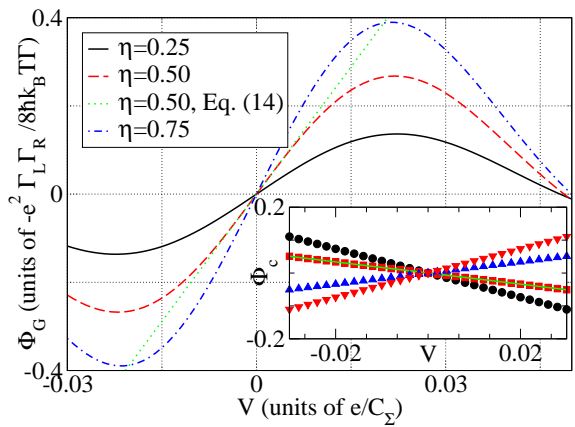}
}
\caption{(Color online). Magnetic-field asymmetry of the
differential conductance versus bias voltage for different
capacitance asymmetries. We set $k_B T=0.01\mathcal{U}$ (${\mathcal U}=e^2/C_\Sigma$),
$C_1+C_2=C_g=0.5$, $\Delta=0.1{\mathcal U}$, $\Gamma=0.002{\mathcal U}$ and $V_g=1.173 e/C_\Sigma$, which
corresponds to a maximum in Fig.~\ref{fig1}. Dotted line
shows the low voltage result given by Eq.~(\ref{eq_phig}).
Inset: Cotunneling magnetic-field asymmetry (in units of $\Gamma_L\Gamma_R/{\mathcal U}^2$)
versus bias for $e V_g/{\mathcal U}=1.8,1.9,2.1,2.2$ (on the left, from top to bottom).
Full line is obtained from Eq.~(\ref{eq_phic}) for $e V_g/{\mathcal U}=1.9$.}
\label{fig2}
\end{figure}

%Eq. (11) determines $\Phi_G$. 
We illustrate the behavior of $\Phi_G$
in Figs.~\ref{fig1} and~\ref{fig2}.
We define a capacitance asymmetry factor,
\begin{equation}
\eta=\frac{C_1-C_2}{C_1+C_2}\,.
\end{equation}
Clearly, $\Phi_G$ is nonzero only for asymmetric couplings.
In Fig.~\ref{fig1}, we show $\Phi_G$ as a function of the back-gate
voltage $V_g$ for a finite bias and different temperatures.
For simplicity, we set $E_F=0$ and take uniformly spaced
levels: $\Delta=\varepsilon_N-\varepsilon_{N-1}$
independent of $N$ (in reality,
levels are Wigner-Dyson distributed).
The curve is periodic since $\Phi_G$ reflects
the periodicity of the conductance. 
The asymmetry vanishes exactly at the degeneracy points,
i.e., at gate voltages $V_g=e(N-1/2)/C_g+\varepsilon_N C_\Sigma/e C_g$
(or simply $V_g=e(N-1/2)/C_g$ for $\Delta\ll \mathcal{U}$),
where the conductance is maximum as $\tilde{\varepsilon}_N=0$.
Importantly, $|\Phi_G|$ reaches the maximum value on both
sides of the degeneracy point and then decreases in the
Coulomb-blockade valley, where the charge is fixed,
because no transport is permitted. 
For very low voltages ($eV\ll k_B T$) a compact analytic expression can be found:
\begin{eqnarray}\label{eq_phig}
\Phi_G&=&-\frac{e}{\hbar} \frac{\Gamma_L^N \Gamma_R^N}{4\Gamma^Nk_B T}
\frac{\eta eV}{k_BT} \\ \nonumber
&\times&\frac{C_\Sigma-C_g}{C_\Sigma}
\cosh^{-2} \frac{\tilde{\varepsilon}_N}{2k_BT}
\tanh \frac{\tilde{\varepsilon}_N}{2k_BT}\,.
\end{eqnarray}
We find that the maxima of $|\Phi_G|$ take place approximately at
$\tilde{\varepsilon}_N=k_B T$, i.e., for gate voltages of the order
of $k_B T$ away from the degeneracy point.
This explains as well why the maxima (minima) shift to
lower (higher) values of $V_g$ with increasing $T$.
Moreover, it is worthwhile to note that the asymmetry
effect vanishes overall in the high-$T$ regime.
This implies that when temperature is higher than the
interaction $e^2/C_\Sigma$ transport is mediated by
thermal fluctuations only, which are $B$-symmetric.
We note in passing that our results are formally
related to the voltage asymmetry that arises in a quantum dot
which is more coupled to, say, the left lead than to
the right lead.\cite{sta96}. As a consequence, the conductance measured
at forward bias differs from the backward bias case.\cite{fox93}.

Figure~\ref{fig2} presents the nonequilibrium conductance
as a function of the bias voltage at a fixed $V_g$ corresponding
to one maximum in Fig.~\ref{fig1}. The asymmetry increases rapidly
with voltage and this increase is sharper for increasing
capacitance asymmetry.

In Ref.~\onlinecite{san04} we distinguished between capacitive asymmetry
and scattering asymmetry, the latter arising from asymmetric
tunnel couplings $\Gamma_L^N \neq\Gamma_R^N$.
Both asymmetries can be varied independently by changing
the height and width of the tunnel barrier separating the dot and the edge states.
This distinction was possible
because
the problem could be solved exactly at all orders in the coupling $\Gamma^N$
(coherent tunneling). When the dot is Coulomb-blockaded, 
tunneling is sequential and tunnel couplings are treated to first order
($\Gamma^N$ is the lowest energy scale).
Thus, the effect of a tunnel asymmetry is trivially
incorporated in our equations since
$\Gamma_L^N \Gamma_R^N/\Gamma^N=(1-\xi^2)/4\Gamma^N$
with the scattering asymmetry factor $\xi=(\Gamma_L^N-\Gamma_R)/\Gamma^N$.
However, in the classical treatment of Coulomb blockade given here, the asymmetry $\Phi_G$ vanishes when $\eta=0$ independently of $\xi$.
To include quantum fluctuations is a difficult task since the charge
$Q_d$ is not simply $Ne$ and the self-consistent procedure
to find the dot potential becomes involved. 
In the absence of 
Coulomb blockade effects, but in the presence of a Hartree potential, 
the task can be solved\cite{chr96}
to all orders in $\Gamma$.

Cotunneling processes contribute to
the conductance to order $\Gamma^2$. Thereby a residual asymmetry is expected
around the conductance minima~\cite{ave90}. We consider elastic cotunneling,
which is the dominant off-resonance mechanism at low bias
when $k_B T\ll\Gamma$, as experimentally demonstrated~\cite{def01}.
Elastic cotunneling consists of the virtual tunneling of an electron
in a coherent fashion without leaving the dot in an excited state.
Hence, our theory for transport between ground states is applicable. 
For definiteness, we investigate the minimum between
the $N=1$ and $N=2$ resonances. Due to large denominators in
Eq.~(\ref{eq_s}) we can use $T=0$ Fermi functions in Eq.~(\ref{eq_cur})
and expand in powers of $\Gamma$. The resulting conductance
goes as $(\Gamma/\mathcal{U})^2$. In the inset of Fig.~\ref{fig2}
we plot the numerical result of the asymmetry of the cotunneling
conductance, $\Phi_c$, as a function of $V$ for gates voltages
around the conductance minimum, which represents the electron-hole (e-h)
symmetry point. Interestingly enough, $\Phi_c$ changes
sign about the minimum and exactly vanishes (not shown) at the
e-h symmetry point since charge fluctuations are quenched there
(the mean charge is $1/2$ per channel).
For $E_F=\varepsilon_1+\Delta/2$ the $G_0$ minimum takes place
at $Q_g=C_g V_g=+e$. Then, to leading order in $(Q_g/e-1)$
we find
\begin{equation}\label{eq_phic}
\Phi_c=-\frac{e^2}{h}\Gamma_L\Gamma_R \frac{192\eta(C_\Sigma-C_g)\mathcal{U} eV}
{C_\Sigma (\Delta+\mathcal{U})^4} \left( \frac{Q_g}{e}-1 \right)\,,
\end{equation}
valid in the limit $eV\ll U$ and $k_B T\ll \Gamma\ll\Delta<U$.
This expression reproduces the effects discussed above and
is in remarkable agreement with the numerical results
(see inset of Fig.~\ref{fig2}).

Thus far we have neglected the spin degeneracy.
When $T$ is further lowered, spin-flip cotunneling processes
lead to Kondo effects and the corrections of the conductance become
of the order of $e^2/h$. Notably, a dependence
on the bias polarity\cite{sim99} due to asymmetric couplings\cite{kra02}
has been observed. 
Therefore, one might expect a large magnetic-field asymmetry.
However, recent works\cite{cho04,cor04} have emphasized the robustness of the e-h
symmetry point in the Kondo regime against external disturbances
which would suggest that also the magnetic field asymmetry vanishes at this point.

{\em Conclusions}.---We have demonstrated that careful consideration of the
interaction between a quantum dot and the edge states to which it is coupled
leads to an out-of-equilibrium charging which is asymmetric under magnetic-field
reversal. Crucial to this result is the chirality of 
the polarization charge. Obviously, any model generating an uneven polarization charge
would similarly and quite generally predict an asymmetry.
Importantly, the temperature scale of the magnetic field 
asymmetry we find is determined by the Coulomb charging 
energy. Consequently, the effect reported here should be readily observable 
in a wide range of systems.  

We thank R. L\'opez for useful discussions
and H. Bouchiat, H. Linke and K. Ensslin
for private communications.
This work was supported by the RTN No. HPRN-CT-2000-00144,
the Spanish contract No. PRIB-2004-9765,
the program ``Ram\'on y Cajal'', the Swiss NSF and MaNEP.

\end{document}